\documentclass[final,1p,times]{elsarticle} 
\usepackage{amssymb} 
\usepackage{amsthm} 
\usepackage{lineno} 

\journal{Nuclear Physics A} 

\usepackage{graphics,epsfig}
\usepackage{color}

\begin{document}

\begin{frontmatter} 

\title{System-size scan of dihadron azimuthal correlations in
 ultra-relativistic heavy ion collisions}

\author{S. Zhang$^{a}$}
\author{Y. H. Zhu$^{a,b}$}
\author{G. L. Ma$^{a}$}
\author{Y. G. Ma$^{a}$\corref{cor1}}
\author{X. Z. Cai$^{a}$}
\author{J. H. Chen$^{a}$}
\author{C. Zhong$^{a}$}
\cortext[cor1]{Corresponding author. Email address:
ygma@sinap.ac.cn (Y. G. Ma)}
\address{$^a$Shanghai Institute of Applied Physics, Chinese Academy of Sciences, Shanghai 201800, China}
\address{$^b$Graduate School of the Chinese Academy of Sciences, Beijing 100080, China}

\begin{abstract} 
System-size dependence of dihadron azimuthal correlations in
ultra-relativistic heavy ion collision is simulated by a
multi-phase transport model. The structure of correlation
functions and yields of associated particles show clear
participant path-length dependences in collision systems with a
partonic phase. The splitting parameter  and root-mean-square
width of away-side correlation functions increase with collision
system size from $^{14}$N+$^{14}$N to $^{197}$Au+$^{197}$Au
collisions. The double-peak structure of away-side correlation
functions can only be formed in sufficient ``large'' collision
systems under partonic phase. The contrast between the results
with partonic phase and with hadron gas could suggest some hints
to study onset of deconfinment.
\end{abstract} 
\end{frontmatter} 
\linenumbers 


\section{Introduction}
Quantum Chromodynamics (QCD) calculation predicted an exotic
quark-gluon matter in QCD phase diagram~\cite{QCD-phasetran} may
be created  in the early stage of heavy ion collisions at
ultra-relativistic energy~\cite{White-papers}. For mapping the QCD
phase diagram and locating QCD phase boundary and critical point
~\cite{phase-th}, one needs to find a way to vary temperature
($T$) as well as chemical potential  ($\mu_{B}$).  The NA61
collaboration and NA49-future collaboration~\cite{NA61EA}
suggested that it can be achieved via a systematic energy (E) and
system-size (A) ($E-A$) scan.

Jet quenching phenomenon has been theoretically predicted
~\cite{jet-quenching} and experimentally observed~\cite{jet-ex}.
So far, dihadron azimuthal correlations have been demonstrated as a
good method to reconstruct particle and energy distribution
induced by the quenched jet. In experiment,
 a double-peak structure was found on the away side  of dihadron azimuthal
 correlation functions~\cite{sideward-peak1,sideward-peak2,sideward-peak3}
 and the indication of conical emission of charged hadrons was reported by
 the STAR collaboration~\cite{sideward-peak4}. The centrality and transverse
 momentum
 dependences of double-peak structure of away-side  correlation functions
 were experimentally  investigated  by RHIC-BNL~\cite{sideward-peak3,STARpT} and
 theoretically simulated in Ref.~\cite{di-hadron}.

These interesting phenomena have attracted some theorists to
explain the physical mechanisms for the origin of the double-peak
structure. These mechanisms include a  Cherenkov-like gluon
radiation model~\cite{Koch}, medium-induced gluon bremsstrahlung
radiation~\cite{large-angle,opaque-media-radiation}, shock wave
model in hydrodynamic equations~\cite{Casalderrey}, waking the
colored plasma and sonic Mach cones~\cite{Ruppert}, sonic booms
and diffusion wakes in thermal gauge-string
duality~\cite{sonic-booms}, jet deflection~\cite{deflection} and
strong parton cascade mechanism and so on
~\cite{di-hadron,three-hadron,time-evolution,pt-dependence,glma-sqm09}.
 Recently, Gyulassy and his collaborators suggest
that the conical emission can stem from universal flow-driven
mechanism~\cite{Gyulassy-flow-driven}. In
Ref.~\cite{ZXu-Shock-ph}, the shock wave phenomena are discussed
in viscous fluid dynamics and kinetic theory. Renk and
Neufeld~\cite{Renk-MachCone-EL} presented their systematic study
of dependence of the Mach cone signal on the energy deposition
into the medium in linearized hydrodynamics. The path-length
effect on the correlations relative to the reaction plane are
studied, respectively, by Jia et al. in a simple
model~\cite{Jia-path-length} and by Ma et al. in AMPT
model~\cite{Ma-path-length}. In the shock wave
model~\cite{Casalderrey}, the emission angle relative to jet is
calculated to be about 1.23 rad for QGP, 1.11 rad for hadronic gas
and zero for mixed phase.  The gluon radiation mechanism for the
double-peak structure~\cite{Koch} suggests that the more energetic
the jet, the smaller the emission angle. In Ref.~\cite{QCDCP-MC},
it is suggested that the suppression or even disappearance of Mach
cone at the QCD critical point due to the attenuation of the sound
mode. While it remains unclear what the main mechanism for the
emergence of the double peak structure is, all the experimental
and theoretical works suggest that it should depend on the nature
of the hot and dense matter created in the
collisions~\cite{hard-hard-ex,soft-soft-ex}. In this paper, we
study the properties of hot-dense matter produced by different
system size by investigating the system-size dependence of
dihadron azimuthal correlations.

In this paper, we present participant path-length, defined as
$\nu$ = 2$N_{bin}$/$N_{part}$~\cite{nu-define} ($N_{bin}$ and
$N_{part}$ are the number of binary collision and participants,
respectively), dependence of the double-peak structure of
away-side  correlation function in the most central collisions
(0-10\%).  The structure of  away-side  correlation function
changes near $^{40}$Ca + $^{40}$Ca collisions at $\sqrt{s_{NN}}$ =
200 GeV in central collisions (0-10\%). The results show obvious
degree of freedom dependence in the system with a partonic phase
or with a pure hadron gas~\cite{Song}, which implies information
of the onset of deconfinement.

\section{Model and analysis method}

In this work, a multi-phase transport model (AMPT)  ~\cite{AMPT},
which is a hybrid dynamic  model, is employed to study dihadron
azimuthal correlations. It includes four main components to
describe the physics in relativistic heavy ion collisions: 1) the
initial conditions from HIJING model~\cite{HIJING}, 2) partonic
interactions modeled by a Parton Cascade model (ZPC)~\cite{ZPC},
3) hadronization (discussed later), 4) hadronic rescattering
simulated by A Relativistic Transport (ART) model~\cite{ART}.
Excited strings from HIJING are melted into partons in the AMPT
version with string melting mechanism~\cite{SAMPT} (abbr.
{`\it{the Melt AMPT version}'}) and a simple quark coalescence
model is used to combine the partons into hadrons. In the default
version of AMPT model~\cite{DAMPT} (abbr. {`\it{the Default AMPT
version}'}), minijet partons are recombined with their parent
strings when they stop interactions and the resulting strings are
converted to hadrons via the Lund string fragmentation
model~\cite{Lund}. The Melt AMPT version undergoes a partonic
phase, while a pure hadron gas is in the Default AMPT version.
Details of the AMPT model can be found in a review
paper~\cite{AMPT} and previous works~\cite{AMPT,SAMPT,Jinhui}.

The analysis method for dihadron azimuthal correlations is similar
to that used in previous experiments
~\cite{soft-soft-ex,sideward-peak2}, which describes the azimuthal
correlation between a high $p_{T}$ particle (trigger particle) and
low $p_{T}$ particles (associated particles). The raw signal can
be obtained by accumulating pairs of trigger and associated
particles into $\Delta\phi = \phi_{assoc} - \phi_{trig}$
distributions in the same event. The background which is expected
mainly from elliptic flow is simulated by mixing event
method~\cite{soft-soft-ex,sideward-peak2}. To reconstruct the
background, we accumulate pairs of one fake trigger particle (high
$p_T$) in one event and another fake  associated particle (low
$p_T$) in another event to obtain the $\Delta\phi$ distribution as
the corresponding background, the centralities of the above two
events are requested very closed. Then the background is
subtracted from raw signal by using ``A Zero Yield At Minimum"
(ZYAM) assumption as that used in experimental
analysis~\cite{sideward-peak2} (see our detailed analysis in
Ref.~\cite{di-hadron}). Recently,  Wang et al.
~\cite{Fuqiang-bg-cluster} discussed the background in the
correlations and presented an analytical form for flow background
to jet-like azimuthal correlations in a cluster approach. And it
is suggested that the collision geometry fluctuations and
triangular flow should be taken into account in the correlation
analysis~\cite{Alver-bg-v3,Ko10,GLMa-v3}. But those go beyond our
discussion in this paper.

\section{Results and discussions}

\subsection{Structure of dihadron correlation function}

The participant path-length, defined as $\nu$ = 2$N_{bin}$/
$N_{part}$  ~\cite{nu-define},  can describe degree of multiple
collisions between participants in the early stage of heavy ion
collisions and characterize the size of the reaction zone. The
$n_{col}^{parton}$  represents average collision number of partons
in the Melt AMPT version. The values of $\nu$ and
$n_{col}^{parton}$ significantly increase with varying collision
system ($CSYS$) from ``small'' size to ``large'' size at
$\sqrt{s_{NN}}$ = 200 GeV in the most central collisions (0-10\%)
as shown in Table~\ref{talble-samecentral}. From this table, we
can see the multiple collisions are more frequent in ``large''
size collision system than in ``small'' size one. The values of
$N_{part}$, $N_{bin}$ and $\nu$ are comparable to those from the
Glauber Model~\cite{sys-glauber} for both $^{64}$Cu + $^{64}$Cu
and $^{197}$Au + $^{197}$Au collisions.

\begin{figure}[htb]
\vspace{-4cm}
\includegraphics[scale=0.55,bb=-80 -35 240 460]{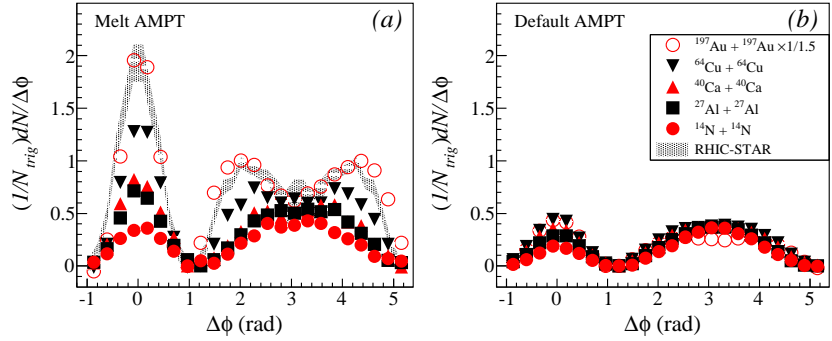}
\vspace{-0.8cm} \caption{Dihadron azimuthal correlation functions
for different collision systems in centrality 0-10\% at
$\sqrt{s_{NN}}$ = 200  GeV; Kinetic windows: $1 < p_{T}^{asso} <
3$ GeV/\textit{c}, $2.5 < p_{T}^{trig} < 6$ GeV/\textit{c},
$|\eta| < $1; the shadowing area from the STAR
data~\cite{sideward-peak1}.\label{corr-fun}} \vspace{-0.5cm}
\end{figure}

When the mixed background which mainly stems from elliptic flow is
subtracted from the raw dihadron correlation signal taken in the
same events, we can get the correlation function.
Fig.~\ref{corr-fun} shows dihadron azimuthal correlation functions
of different collision systems in the most central (0-10\%)
collisions at $\sqrt{s_{NN}}$ = 200 GeV. The correlation functions
are calculated in the kinetic windows, 1 $< p_{T}^{assoc} <$ 3
GeV/$c$ as well as 2.5 $< p_{T}^{trig} <$ 6 GeV/$c$ and $|\eta| <$
1. It shows that the structure of away-side correlation function
changes from the Gaussian-like distribution to double-peak
structure near $^{40}$Ca + $^{40}$Ca collisions with varying
collision system from $^{14}$N + $^{14}$N to $^{197}$Au +
$^{197}$Au collisions in the Melt AMPT version. In this figure,
the amplitude of the correlation function becomes higher with the
increasing of collision system size. The associated particles in
the Melt AMPT version are more abundant than those in the Default
AMPT version.

\begin{table}[htb]
\begin{center}
\fontsize{7pt}{1\baselineskip}\selectfont
\caption{ $N_{part}(CSYS)$, $N_{bin}(CSYS)$, $\nu(CSYS)$ =
  $\frac{2N_{bin}(CSYS)}{N_{part}(CSYS)}$, $n_{col}^{parton}$ in different
    collision system at $\sqrt{s_{NN}}$ = 200 GeV for centrality 0-10 \%, the
    value in blanket is taken from Glauber Model~\cite{sys-glauber}.}
\begin{tabular}{|c|c|c|c|c|c|c|c|}\hline
$CSYS$ & $^{14}$N + $^{14}$N & $^{16}$O + $^{16}$O & $^{23}$Na +
$^{23}$Na & $^{27}$Al + $^{27}$Al & $^{40}$Ca + $^{40}$Ca &
$^{64}$Cu + $^{64}$Cu & $^{197}$Au + $^{197}$Au \\\hline
$N_{part}(CSYS)$ & 20.78 & 24.25 & 35.92 & 43.61 & 65.97 & 107.04
(99.0) & 343.32 (325.9)\\\hline $N_{bin}(CSYS)$ & 19.63 & 23.69
& 41.01 & 54.34 & 91.15 & 179.98 (188.8) & 914.71
(939.4)\\\hline $\nu(CSYS)$ & 1.89 & 1.95 & 2.28 &
2.49 & 2.76 & 3.36 (3.8) & 5.33 (5.7)\\\hline $n_{col}^{parton}$
& 1.31 & 1.44 & 1.93 & 2.23 & 2.79 & 3.80 & 7.24
\\\hline
\end{tabular}

\label{talble-samecentral}
\end{center}
\end{table}

The Default AMPT version is used to compare the properties of the
double-peak structure in partonic phase and in hadron gas. For
investigating the properties of collision system-size dependences
of away-side dihadron azimuthal correlations, we extract the
associated particle yield $N_{away}^{assoc}$, splitting parameter
($D$) (half distance between double peaks on the away side) and
Root Mean Square Width ($\Delta \phi_{rms}$) of away-side
associated particles, which will be discussed in the following
sections, respectively.

\subsection{Yield of associated particles}
\begin{figure}[htbp]
\vspace{-1cm}
\includegraphics[scale=0.54,bb=-80 -35 240 460]{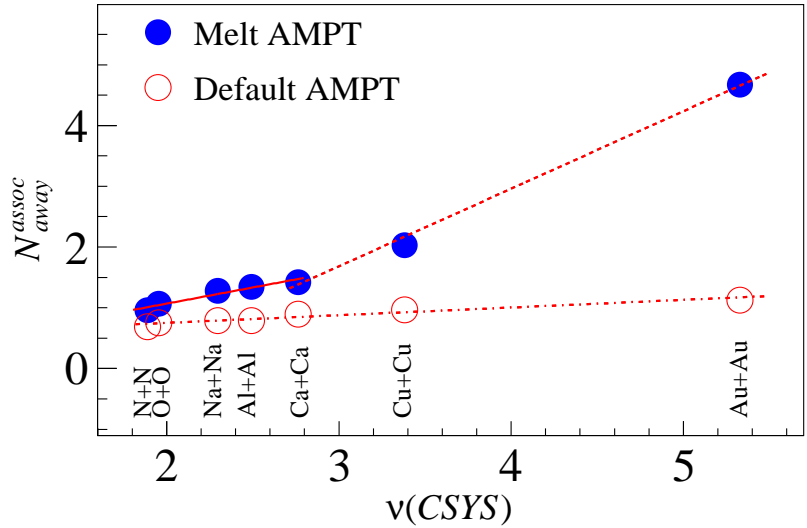}
\vspace{-0.55cm} \caption{ Yield of associated particles on
away-side correlation functions, $N_{away}^{assoc}$, as a function
of $\nu(CSYS)$ by the Melt/Default AMPT version in the centrality
of 0-10\% at $\sqrt{s_{NN}}$ = 200 GeV.\label{yield} }
\end{figure}

The $\nu(CSYS)$ dependence of $N_{away}^{assoc}$ is shown in
Fig.~\ref{yield}  from the Melt/ Default AMPT version,
respectively. It presents a significant increasing trend of
$N_{away}^{assoc}$ with varying the collision system from $^{14}$N
+ $^{14}$N to $^{197}$Au + $^{197}$Au collisions in the most
central collisions (0-10\%) at $\sqrt{s_{NN}}$ = 200 GeV in the
Melt AMPT version. The Default AMPT version, with a hadronic gas,
does not result in a rapid increasing dependence trend.  In the
Melt AMPT version, the dependence trend indicates the jet
correlation information can be inherited by and transmitted to
more particles in a partonic phase than in a hadronic gas,
especially in ``large'' size collision system. Furthermore it
implies that the interaction strength in ``large'' size collision
system is more significant than that in ``small'' size collision
system, and while strong parton cascade plays a dominant role to
push more away-side associated  particles in the Melt AMPT
version. It is interesting that the increasing slope of
$N_{away}^{assoc}$ vs $\nu(CSYS)$  from the linear fitting in the
Melt AMPT version is quicker above $^{40}$Ca + $^{40}$Ca collision
system, where clear double-peak structure emerges, than that in
small systems. This property indicates the double-peak (Mach-like)
structure can enhance associated particles yields of jet
correlations partially.

\subsection{$\Delta \phi_{rms}$ and splitting parameter on the away side }

\begin{figure}[htb]
 \vspace{-1cm}
\includegraphics[scale=0.54,bb=-80 -35 240 460]{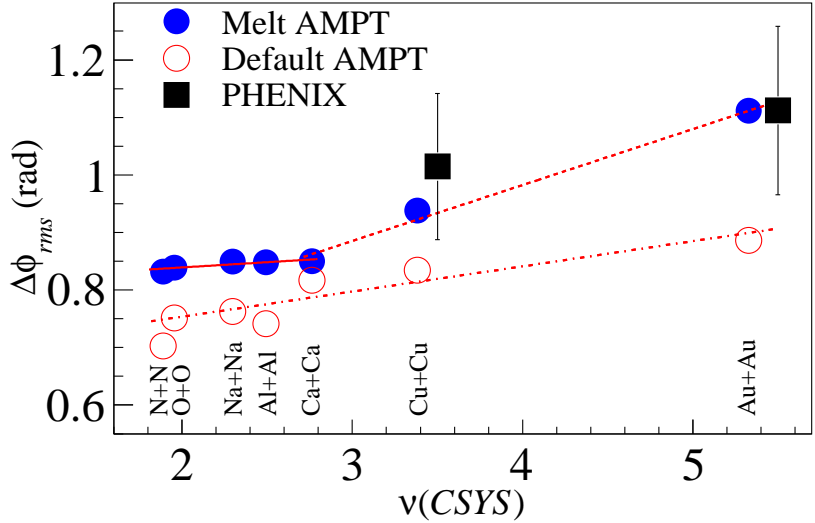}
 \vspace{-0.55cm}
\caption{RMS-Width ($\Delta \phi_{rms}$)  of the away-side
correlation functions as a function of $\nu(CSYS)$ in the
Melt/Defualt AMPT version for the centrality 0-10\% at
$\sqrt{s_{NN}}$ = 200 GeV; Square from PHENIX
data~\cite{sideward-peak2,rmsdata}.}\label{rms}
\end{figure}

Root Mean Square Width ($\Delta \phi_{rms}$) of away-side
correlation function is defined as
\begin{displaymath}
\Delta
\phi_{rms}=\sqrt{\frac{\sum\limits_{away}{(\Delta\phi-\Delta\phi_{m})^{2}(1/N_{trig})(dN/d\Delta\phi)}}{\sum\limits_{away}{(1/N_{trig})(dN/d\Delta\phi)}}},
\end{displaymath}
where $\Delta\phi_{m}$ is the mean $\Delta\phi$ of away-side
correlation function and  it approximates to $\pi$. $\Delta
\phi_{rms}$ can describe the diffusion degree of the associated
particles relative to the direction of back jet. The $\nu(CSYS)$
dependences of $\Delta \phi_{rms}$ in the Melt/ Default AMPT
version are shown in Fig.~\ref{rms}, respectively. $\Delta
\phi_{rms}$ from the Melt AMPT version are consistent with PHENIX
data~\cite{sideward-peak2,rmsdata} for Cu + Cu and Au + Au
collisions. $\Delta\phi_{rms}$ increases from $^{14}$N + $^{14}$N
collisions to $^{197}$Au + $^{197}$Au collisions in the Melt AMPT
version as well as the Default AMPT version, but the increasing
trend is not so quick in the later, especially for systems larger
than $^{40}$Ca + $^{40}$Ca. The increasing trend of $\Delta
\phi_{rms}$ shows broadening of away-side correlation functions
with increasing size of collision system. It indicates that the
jet correlation information can reach faraway relative to
direction of jet with changing the collision system from ``small''
size one to ``large'' size in a partonic phase. It is remarkable
that the increasing trend of $\Delta \phi_{rms}$ from the linear
fitting in the Melt AMPT version shows two different slope after
and before  $^{40}$Ca + $^{40}$Ca collision system, where clear
double-peak structure emerges.  The double peak (Mach-like)
phenomenon from quenched jet can enhance the diffusion degree of
the associated particles relative to back jet in the ``large''
size collision system. This suggests that the back jet
modification in the medium created in heavy ion collisions with a
partonic phase is more distinct in the ``large'' size collision
system than in ``small'' size one.

\begin{figure}[htb]
 \vspace{-1cm}
\includegraphics[scale=0.54,bb=-80 -35 240 460]{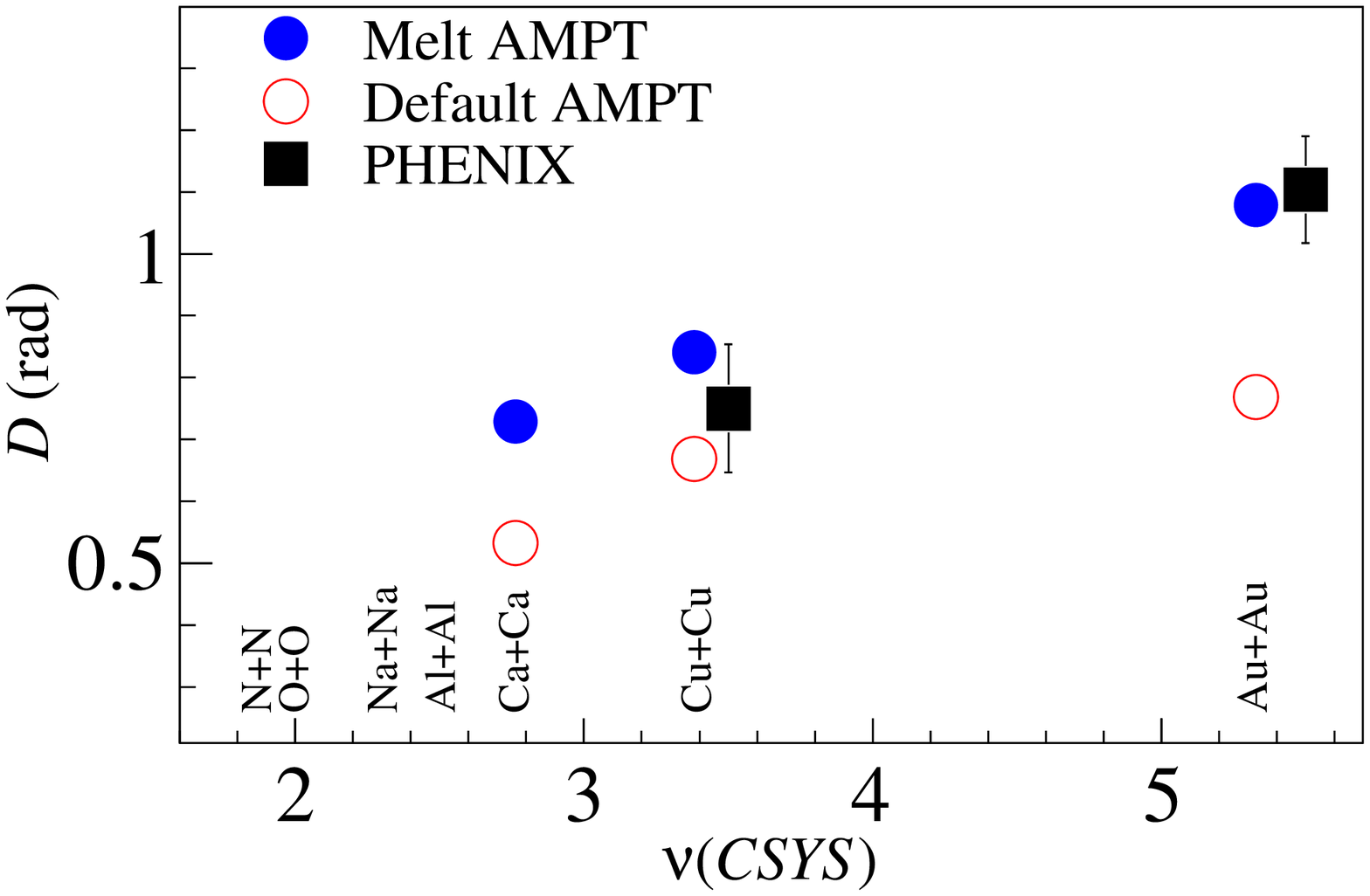}
\vspace{-0.55cm} \caption{ Splitting parameter ($D$) of the
away-side correlation functions as a function of $\nu(CSYS)$ in
the Melt/Defualt AMPT version for the centrality 0-10\% at
$\sqrt{s_{NN}}$ = 200 GeV; Square from PHENIX
data~\cite{sideward-peak3}.\label{splitd}}
\end{figure}

The splitting parameter ($D$) is another useful observable to
characterize the structure of  the double-peak of away-side
correlation function, and further discloses essential of jet
modification. The $\nu(CSYS)$ dependence of splitting parameters
($D$) in the Melt/Default AMPT version are shown in
Fig.~\ref{splitd} for the most central collisions at
$\sqrt{s_{NN}}$ = 200 GeV. For lighter systems from $^{14}$N +
$^{14}$N to  $^{27}$Al + $^{27}$Al collisions, the splitting
parameter is not extracted  since there is no observable
double-peak structure of away-side correlation functions. In both
simulation cases, the splitting parameter ($D$) increases from
``small'' size collision system to ``large'' size one, which
indicates that there exists  stronger jet-medium interaction in
``large''  system. It is also remarkable that the splitting
parameter ($D$) is larger in the Melt AMPT version than that in
the Default AMPT version. The Melt AMPT results are comparable to
PHENIX data~\cite{sideward-peak3} for Cu + Cu and Au + Au
collisions due to effect of parton cascade in the Melt AMPT
version~\cite{di-hadron}. The parton interaction cross section is
taken to be 10 mb in this work, which is also reasonable for
reproducing elliptic flow as well as  dihadron azimuthal
correlations in the Melt AMPT
version~\cite{AMPT,SAMPT,Jinhui,di-hadron}. This calculation
reflects that it is necessary to pass a strong partonic stage to
reproduce large enough double-peak structure as the experimental
data demonstrate.

From these results, it can be concluded that a considerable
``large'' collision  system is necessary and the strong parton
cascade is essential for the formation of the double-peak
structure of away-side  correlation function. An onset of the
observable double-peak structure occurs in the  mass range of
$^{40}$Ca + $^{40}$Ca collision. This phenomenon indicates that
the correlation is sensitive to $\nu(CSYS)$ and
$n_{col}^{parton}$, i.e. the correlation depends on the collision
system size and the violent degree of the partonic interaction in
a partonic phase. Different results obtained in a partonic phase
and a pure hadron gas imply the double-peak structure and jet
modification  are sensitive to the effective degree of freedom of
the dense medium created in relativistic heavy ion collisions,
which can give us some hints of the onset of deconfinement in the
system-size viewpoint.

\section{Summary}
In summary, the present work discusses the collision system-size
dependence of  dihadron azimuthal correlations at $\sqrt{s_{NN}}$
= 200 GeV by a multi-phase transport model. The yields of
associated particles, width of away-side correlation functions and
splitting parameter show significant system-size dependence. The
away-side correlation function becomes more broadening with the
increasing of collision system size and displays the  onset of
double-peak structure near $^{40}$Ca + $^{40}$Ca collisions after
the system undergoes a strong partonic transport stage. These
results also present the degree of freedom dependence, which might
be related to onset of deconfinement. We would remark that these
observations do not assume any dynamical mechanism for the
formation of the double hump structure. The AMPT model would
include collective Mach-like effects associated to the particles,
but it would also include eccentricity fluctuating effects and
triangle flow components. The result is, in this sense, robust.

\section*{Acknowledgements}
This work was supported in part by  the National Natural Science
Foundation of China under Grant No. 11035009, 11047116, 10905085,
and 10875159, 10705043 and 10705044, and the Shanghai Development
Foundation for Science and Technology under contract No.
09JC1416800, and the Knowledge Innovation Project of the Chinese
Academy of Sciences under Grant No. KJCX2-EW-N01, Y155017011 and
O95501P0-11 and the Project-sponsored by SRF for ROCS, SEM.
O819011012.


\end{document}